\documentclass[sigconf, nonacm, preprint]{acmart}
\usepackage[table]{xcolor}
\usepackage{graphicx}  
\usepackage{xcolor}    
\usepackage{colortbl}  
\usepackage{booktabs}  
\usepackage{listings}  
\usepackage[table]{xcolor} 
\usepackage{balance}
\AtBeginDocument{%
  }

\acmISBN{978-1-4503-XXXX-X/2018/06}

\begin{document}

\title{Transparent and Controllable Recommendation Filtering via Multimodal Multi-Agent Collaboration}

\author{Chi Zhang}
\authornote{Both authors contributed equally to this research.}
\email{zhangchi23@m.fudan.edu.cn}
\affiliation{%
  \institution{Fudan University}
  \city{Shanghai}
  \country{China}
}

\author{Zhipeng Xu}
\authornotemark[1] 
\email{zpxu23@m.fudan.edu.cn}
\affiliation{%
  \institution{Fudan University}
  \city{Shanghai}
  \country{China}
}

\author{Jiahao Liu}
\email{jiahaoliu23@m.fudan.edu.cn}
\affiliation{%
  \institution{Fudan University}
  \city{Shanghai}
  \country{China}
}

\author{Dongsheng Li}
\email{dongshengli@fudan.edu.cn}
\affiliation{%
  \institution{Microsoft Research Asia}
  \city{Shanghai} 
  \country{China}
}

\author{Hansu Gu}
\email{hansug@acm.org}
\affiliation{
  \institution{Independent}
  \city{Seattle}
  \country{United States}
}

\author{Peng Zhang}
\authornote{Corresponding author.} 
\email{zhangpeng_@fudan.edu.cn}
\affiliation{%
  \institution{Fudan University}
  \city{Shanghai}
  \country{China}
}

\author{Ning Gu}
\email{ninggu@fudan.edu.cn}
\affiliation{%
  \institution{Fudan University}
  \city{Shanghai}
  \country{China}
}

\author{Tun Lu}
\authornotemark[2]
\email{lutun@fudan.edu.cn}
\affiliation{%
  \institution{Fudan University}
  \city{Shanghai}
  \country{China}
}

\renewcommand{\shortauthors}{Zhang and Xu, et al.}

\begin{abstract}
While personalized recommender systems excel at content discovery, they frequently expose users to undesirable or discomforting information, highlighting the critical need for user-centric filtering tools. Current methods leveraging Large Language Models (LLMs) struggle with two major bottlenecks: they lack multimodal awareness to identify visually inappropriate content, and they are highly prone to "over-association"---incorrectly generalizing a user's specific dislike (e.g., anxiety-inducing marketing) to block benign, educational materials. These unconstrained hallucinations lead to a high volume of false positives, ultimately undermining user agency. To overcome these challenges, we introduce a novel framework that integrates end-to-cloud collaboration, multimodal perception, and multi-agent orchestration. Our system employs a fact-grounded adjudication pipeline to eliminate inferential hallucinations. Furthermore, it constructs a dynamic, two-tier preference graph that allows for explicit, human-in-the-loop modifications (via $\Delta$-adjustments), explicitly preventing the algorithm from catastrophically forgetting fine-grained user intents. Evaluated on an adversarial dataset comprising 473 highly confusing samples, the proposed architecture effectively curbed over-association, decreasing the false positive rate by 74.3\% and achieving nearly twice the F1-Score of traditional text-only baselines. Additionally, a 7-day longitudinal field study with 19 participants demonstrated robust intent alignment and enhanced governance efficiency. User feedback confirmed that the framework drastically improves algorithmic transparency, rebuilds user control, and alleviates the fear of missing out (FOMO), paving the way for transparent human-AI co-governance in personalized feeds.
\end{abstract}





\maketitle

\section{Introduction}
\begin{figure*}[t]
  \centering
  \includegraphics[width=0.96\textwidth]{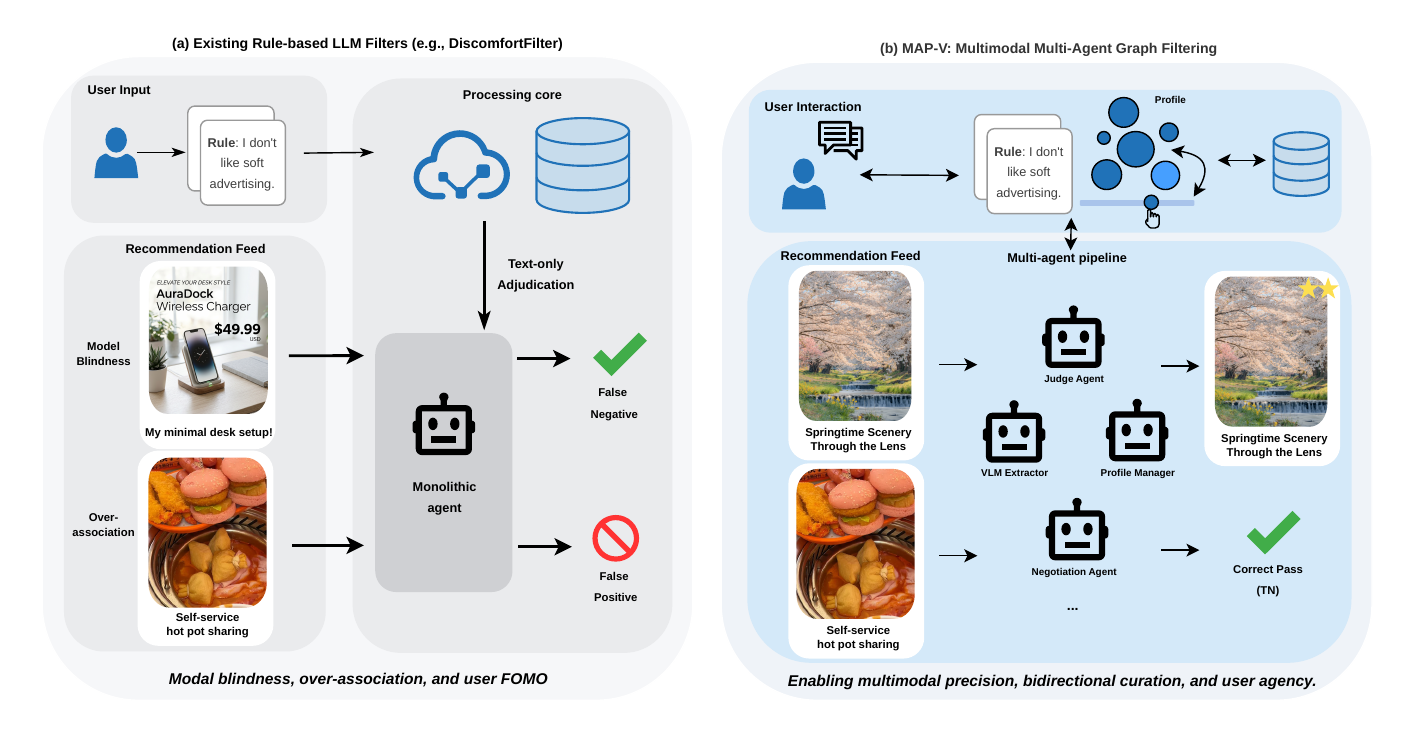} 
  \vspace{-5mm}
  \caption{Comparison of recommendation filtering paradigms. (a) Existing text-only monolithic filters often suffer from \textit{modal blindness} (failing to detect visual-only ads) and \textit{over-association} (misclassifying benign content). (b) MAP-V leverages a multimodal multi-agent pipeline and an editable preference graph to achieve precise intent alignment, enabling both accurate factual adjudication and proactive content curation (via Star Badges).}
  \label{fig:teaser}
  \vspace{-3mm}
\end{figure*}

Personalized recommender systems have become the fundamental infrastructure for information distribution across modern web platforms\cite{covington2016deep, zhang2019deep}. By analyzing historical user behaviors, these algorithms deliver tailored content that significantly enhances user engagement and satisfaction. However, the ubiquitous optimization for click-through rates often inadvertently exposes users to ``discomforting recommendations.''\cite{stray2021what} Content that triggers discomfort is highly subjective and context-dependent: a user experiencing academic anxiety might feel distressed by targeted advertisements for graduate school tutoring, while another might perceive ``clickbait'' covers as offensive. To mitigate such negative experiences, platforms typically provide native feedback mechanisms, such as ``Not Interested'' buttons. Unfortunately, these mechanisms operate as unidirectional black boxes. Users remain oblivious to how their feedback shapes the underlying profiling model and are often deterred from aggressively filtering content due to the Fear Of Missing Out (FOMO) on genuinely valuable information. Consequently, users are stripped of their algorithmic agency, relegated to passive consumers rather than active curators of their digital environments.

Recent advancements have explored the integration of Large Language Models (LLMs) to empower user controllability. A pioneering work, DiscomfortFilter, leverages LLMs to translate natural language feedback into editable filtering rules, demonstrating the feasibility of interactive content moderation (as contrasted in Figure \ref{fig:teaser}a). Despite improving user agency, such text-only, monolithic LLM approaches exhibit two critical limitations when deployed in complex, real-world recommendation streams. First, they suffer from a profound \textbf{semantic gap in multimodality}. Real-world feeds on image-centric social platforms are inherently multimodal. A text-only LLM is effectively blind to ``image-text mismatch'' violations---where benign titles deliberately conceal inappropriate, clickbait, or implicitly promotional visual content---leading to a high rate of False Negatives (FN). Second, they are prone to \textbf{severe over-association}. When handling highly subjective human preferences, a monolithic LLM acting simultaneously as intent parser and judge tends to over-generalize. For instance, if a user specifies a dislike for ``the anxiety caused by technology promotion,'' the model might hallucinate and broadly block objective, educational content such as ``the introduction of a technological evolution framework.'' This unconstrained inferential association generates excessive False Positives (FP), rapidly eroding user trust and compounding the system's tendency to unintentionally ``forget'' fine-grained user constraints during continuous profile updating.

To systematically address these challenges, we introduce \textbf{MAP-V (Multimodal Agentic Profiling \& Verification)}, an end-cloud collaborative, multimodal, multi-agent content filtering system. MAP-V shifts the paradigm of recommendation filtering from a black-box shield to a transparent, bidirectional co-governance tool. Specifically, to overcome modal blindness and over-association, MAP-V introduces a decoupled \textit{multimodal multi-agent} adjudication architecture. We separate intent parsing from filtering execution. The Judge Agent strictly adheres to a ``fact-grounded, no speculative association'' prompting principle (Appendix~\ref{app:system_prompts}), leveraging three-layer structured visual features (perception, cognition, and semantics) extracted by Vision-Language Models (VLMs). This is accompanied by a local CLIP-based\cite{radford2021learning} vector microservice for robust fallback during cloud latency. This architecture ensures that filtering decisions are rigorously based on verifiable multimodal facts rather than unwarranted LLM inferences.

Furthermore, to prevent the forgetting of fine-grained intents and rebuild user trust, MAP-V maintains a \textit{dual-layer dynamic preference graph}. On the backend, a Reflection-based Adaptive Hierarchy (RAH) agent utilizes local MiniLM embeddings\cite{wang2020minilm} and PageRank algorithms\cite{page1999pagerank} for long-term semantic generalization. On the frontend, users interact with a visualized behavioral profile where they can directly drag sliders to manipulate feature weights. These manual $\Delta$-adjustments are recorded as explicit biases that dynamically override implicit algorithmic merging, ensuring that human intent maintains absolute priority. Moving beyond passive filtering, MAP-V introduces a bidirectional curation mechanism: it computes preference scores to assign ``Star Badges'' to highly matched content, effectively mitigating user FOMO and enhancing content discovery efficiency (Figure \ref{fig:teaser}b).

We validated the effectiveness of MAP-V through a rigorous mixed-methods evaluation. First, to test the algorithm's upper bounds against over-association and the modal gap, we constructed an offline adversarial benchmark comprising 473 high-confusion samples (e.g., conceptual confusion, emotional nuance, and image-text mismatch). The results show that MAP-V reduced the FP rate by 74.3\% and increased the F1-Score to 0.7143 compared to the text-only baseline, demonstrating that multimodal perception and multi-agent decoupling effectively mitigate inferential hallucinations. Second, we deployed MAP-V in a 7-day in-the-wild longitudinal study ($N=19$).  Objective telemetry and user feedback revealed that by precisely intercepting toxic interactions on the client side, MAP-V substantially reduced users' exposure to low-quality content and minimized the cognitive burden of manual filtering. Subjective evaluations further confirmed that MAP-V’s high accuracy and transparency significantly enhanced user controllability, allowing for a cleaner information environment with far less manual intervention than native platforms.

The main contributions of this work are summarized as follows:
\begin{itemize}
    \item We propose MAP-V, an end-cloud collaborative filtering architecture that utilizes multimodal VLM perception and multi-agent decoupling (Intent Parser, Judge, Dispute, and RAH Agents). This design effectively suppresses the ``over-association'' hallucination and bridges the semantic gap inherent in pure-text LLM filters.
    
    \item We design a dual-layer dynamic preference graph incorporating an explicit human $\Delta$-bias memory mechanism. This innovation resolves the unintended forgetting of fine-grained user intents, transforming black-box filtering into a transparent, user-editable bidirectional curation tool.
    
    \item Through a rigorous offline adversarial benchmark ($N=473$) and a 7-day in-the-wild longitudinal user study, we quantitatively and qualitatively demonstrate MAP-V's superiority. It achieves a near-doubled F1-Score in complex scenarios, significantly restores user algorithmic agency, and fosters a healthier recommendation ecosystem.
\end{itemize}
\vspace{-4mm}
\section{Related Work}
\subsection{Discomforting and Unwanted Content Filtering}
Filtering discomforting or inappropriate content in recommender systems has garnered increasing attention. Traditional content moderation methods heavily rely on extensive keyword blocklists or rule-based heuristics \cite{davidson2017automated}. While efficient, these rigid approaches lack the capacity to comprehend natural language nuances and contextual cues, frequently resulting in massive false positives. As users' perceptions of discomfort are highly subjective, researchers have advocated for personalized moderation \cite{jhaver2023personalizing}. With the advent of Generative AI, leveraging Large Language Models (LLMs) to translate natural language intents into filtering rules (e.g., DiscomfortFilter \cite{liu2025filtering}) has emerged as a promising trend. However, as noted in recent literature \cite{liu2025filtering}, deploying monolithic, closed-source LLMs as real-time filters for high-throughput recommendation streams is widely deemed impractical due to their vulnerability to prompt injection attacks\cite{greshake2023not} and prohibitive inference costs. To circumvent these engineering bottlenecks, existing studies compromise by using LLMs merely to generate static text-based filtering rules. Yet, this compromise leads to a critical algorithmic flaw: these monolithic, text-only systems are prone to inferential hallucinations when processing vague or subjective expressions. Lacking a systematic mechanism to prevent ``over-association,'' they struggle to maintain filtering precision in highly confusing or emotionally nuanced scenarios. MAP-V addresses this gap by introducing an end-cloud collaborative, multi-agent cross-verification architecture, explicitly decoupling intent parsing from content adjudication to strictly suppress inferential hallucinations while ensuring secure, cost-effective real-time execution.
 \vspace{-4mm}
\subsection{Multimodal Understanding for Recommendation}
Multimodal learning has been extensively explored to enhance recommender systems. Classic models, such as MMGCN \cite{wei2019mmgcn} and MKGAT \cite{sun2020multi}, utilize multi-modal graph convolutional networks and knowledge graph attention mechanisms to achieve fine-grained feature alignment. However, the primary objective of these works is to fuse features (e.g., visual, acoustic, and textual) to boost engagement metrics like Click-Through Rate (CTR). In the context of personalized content filtering, multimodal features are not merely an ``enhancing signal'' but an absolute necessity to bridge the semantic gap. To evade text-based moderation, undesirable or borderline content often employs an ``image-text mismatch'' strategy\cite{ha2021automatically}---using benign text to mask violating visual elements. Therefore, constructing a multimodal filtering pipeline capable of simultaneously scrutinizing and inferring from both text and visual facts, much like a human moderator, is imperative. MAP-V pioneers this by integrating Vision-Language Models (VLMs) into the filtering pipeline, ensuring that decisions are grounded in verifiable visual evidence.

\subsection{Human-in-the-Loop and Controllable RecSys}
To counter the trust crisis engendered by black-box recommendation algorithms, researchers have increasingly integrated ``Human-in-the-Loop'' (HITL) principles into system design. Early explorations include critiquing-based mechanisms \cite{chen2012critiquing} that elicit interactive preferences, and explainable AI paradigms \cite{knijnenburg2012explaining} designed to foster user trust. Recently, user-controllable recommender systems \cite{wang2022user} and LLM-powered recommendation agents (e.g., the RAH! framework \cite{shu2024rah}) have been proposed to mitigate filter bubbles by allowing users to converse with the system. While these methods improve interactivity, they face a critical challenge during long-term engagement: users' fine-grained intents are often subjected to ``catastrophic forgetting,''\cite{kirkpatrick2017overcoming} gradually swallowed by complex, automated algorithmic weights. MAP-V establishes a novel academic standpoint to resolve this pain point. By introducing a visualizable preference graph with explicit manual $\Delta$-adjustments, MAP-V effectively decouples the implicit algorithmic evolution from explicit human intervention. This mechanism grants users absolute, transparent, and persistent control over the algorithm, ensuring that human directives consistently override automated generalization.

\section{MAP-V: System Architecture and Preference Modeling}

\begin{figure*}[t]
  \centering
  \includegraphics[width=0.90\textwidth]{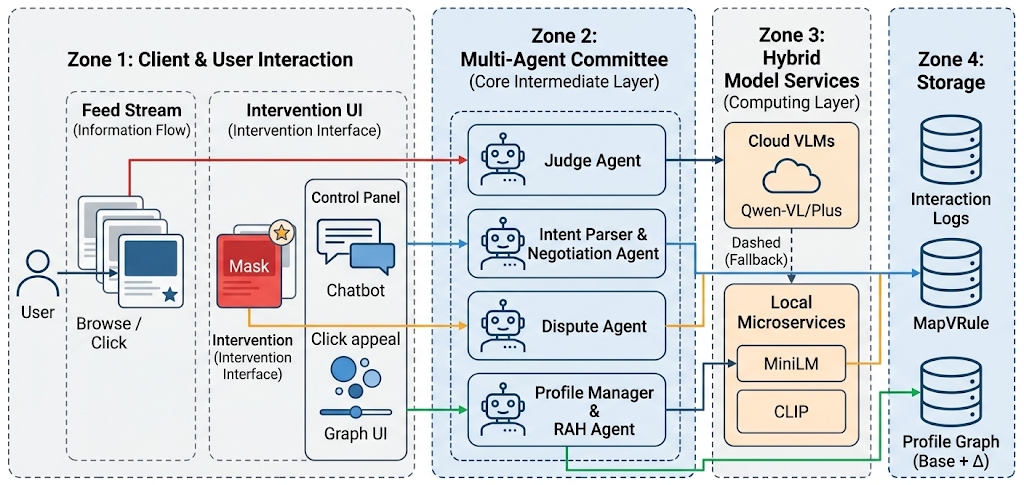}
  \vspace{-2mm}
  \caption{The detailed architecture of the MAP-V system. It contains four functional zones: Client Intervention, Multi-Agent Backend, Hybrid Model Services, and Knowledge Storage. The red path highlights dual-layer filtering adjudication, while blue/green paths denote continuous intent alignment and preference-graph evolution.}
  \label{fig:architecture}
  \vspace{-4mm}
\end{figure*}

\subsection{Design Goals and Problem Formulation}
MAP-V is designed for \textit{real-time and controllable} discomfort filtering in multimodal recommendation feeds. For each incoming item, we denote
\begin{equation}
	x=(t,c,I,\tau),
\end{equation}
where $t$ is title text, $c$ is available snippet text (possibly truncated by platform preview), $I$ is optional image content, and $\tau$ represents auxiliary metadata (e.g., tags). The user maintains an editable rule set $\mathcal{R}$, where each rule is defined as
\begin{equation}
	r=(d,w,m),\quad w\in[-1,1],\; m\in\{\texttt{text},\texttt{image},\texttt{image\_text}\}.
\end{equation}
Here, $d$ is a natural-language description, $w$ is rule weight (negative for filtering, positive for allowance/recommendation), and $m$ is modality scope.

Following a decision-theoretic view, MAP-V can be abstracted as an adjudication function
\begin{equation}
  f(x, R, G_u) \rightarrow \{y_{block}, y_{star}\},
\end{equation}
where $i$ denotes a recommended item (text-image pair), $R_u$ is the user-editable rule repository, and $G_u$ is the dual-layer preference representation.

Given $(x,\mathcal{R})$, MAP-V outputs a binary filtering decision $y_{block}\in\{0,1\}$, an optional triggered-rule identifier, and a fact-grounded rationale. For unblocked items, MAP-V further computes a preference score $y_{star}\in[0,1]$ to render a Star Badge for positive curation.

\subsection{End-Cloud Collaborative Architecture}
MAP-V adopts an end-cloud collaborative architecture with four coordinated zones: (1) Client Intervention Layer, (2) Multi-Agent Backend Control Layer, (3) Hybrid Model Services Layer, and (4) Knowledge \& Storage Layer, as shown in Figure~\ref{fig:architecture}.

\textbf{Client intervention layer.}
The browser extension intercepts feed cards and provides explicit controls, including \textit{Filter}, \textit{Appeal}, and profile \textit{Re-weight}. Unlike black-box recommendation systems, users can directly inspect and intervene in both filtering outcomes and preference-state evolution.

\textbf{Backend control layer.}
The Django ASGI backend orchestrates rule loading, agent invocation, adjudication snapshot logging, and preference graph synchronization. It also maintains auditable traces for each filtered item to support later dispute resolution.

\textbf{Hybrid model services.}
Cloud models provide high-quality multimodal reasoning (Qwen-VL-Plus\cite{bai2023qwenvl} for visual structuring, Qwen-Plus\cite{bai2023qwen} for final judgment), while local services (MiniLM + vector retrieval and CLIP fallback) guarantee low-latency similarity computation and robust degradation under cloud failures.

\textbf{Knowledge and storage layer.}
MAP-V persistently stores user rules, rule versions, profile nodes/edges, interaction records, and visual-analysis caches. This allows long-horizon preference evolution while preserving operation-level interpretability.

\subsection{Cloud Multi-Agent Fact-Grounded Adjudication}
The design motivation is to directly address two recurring failure modes of monolithic LLM moderation: \textit{inferential hallucination} (over-association beyond evidence) and \textit{modal blindness} (insufficient grounding across text-image signals). To mitigate these issues, MAP-V replaces the monolithic paradigm with a narrow-role multi-agent pipeline.\cite{wu2023autogen}

\vspace{1mm}\noindent\textbf{Vision Extractor Agent with structured output.}
When image $I$ exists, Qwen-VL-Plus first generates three-tier visual evidence (Perception/Cognition/Semantics) in constrained JSON form (Appendix~\ref{app:system_prompts}, \textit{Qwen-VL Visual Feature Extraction Prompt}). The agent is only allowed to "see and describe" and is not responsible for rule matching. Results are cached by image-URL MD5 to reduce repeated latency and API costs.

\vspace{1mm}\noindent\textbf{Judge Agent with strict fact grounding.}
Qwen-Plus receives $(t,c,\tau)$, optional visual evidence, and active rules $\mathcal{R}$. The prompt enforces \textit{Fact-grounded, no speculative association}(Appendix~\ref{app:system_prompts}, \textit{Qwen-Plus Rule Judge Prompt}): rule triggering must be supported by explicit textual or visual facts. For truncated snippets, the Judge Agent is instructed not to over-penalize merely because complete body text is unavailable.

\vspace{1mm}\noindent\textbf{Safety Net: Local Cross-Modal Fallback.}
If cloud vision parsing fails, MAP-V switches to local CLIP cross-modal matching between image content and negative-rule descriptions. Once similarity exceeds threshold ($\tau_{clip}=0.30$), the item is conservatively blocked to avoid protection downtime.

\subsection{Interactive Human-in-the-Loop Rule Evolution}
MAP-V supports bidirectional, explicit interaction loops for continuous alignment.

\vspace{1mm}\noindent\textbf{Active Intent Parsing.}
When a user expresses discomfort via natural language, the \textit{Intent Parser Agent} dynamically extracts structured entities, modalities, and emotional weights to formulate a tentative rule. This shifts the paradigm from mechanical keyword blocking to nuanced semantic shielding. A preview-then-confirm protocol is applied, so users can edit the proposal before activation; confirmed updates are versioned and synchronized to the local vector index.

\vspace{1mm}\noindent\textbf{Appeal-Driven Dispute Resolution.}
When an automatic block is deemed a ``False Positive'' by the user (e.g., sarcastic meta-commentary), they can initiate an ``Appeal'' (demonstrated in the UI walkthrough in Appendix \ref{app:ui_details}). The \textit{Dispute Agent} accesses the \textit{Adjudication Dossier} (the context snapshot of the filtering rationale) and negotiates a fine-grained boundary refinement with the user. It autonomously drafts exemption sub-rules (e.g., ``unless the text exhibits a satirical tone'') and operationally returns one actionable proposal (rule modification or allow-rule addition). This closed-loop human-AI negotiation effectively prevents the catastrophic forgetting of nuanced intents without compromising the system's defensive integrity.

\subsection{Dual-Layer Preference Modeling and Star Badge}
Balancing \textit{immediate controllability} and \textit{long-term generalization} is a core challenge in human-in-the-loop recommendation governance. MAP-V resolves this tension by a decoupled dual-layer preference design: the frontend layer absorbs rapid user intent shifts, while the backend layer maintains stable semantic structure against \textit{Catastrophic Forgetting}.

\vspace{1mm}\noindent\textbf{Frontend explicit profile with $\Delta$-bias.}
The frontend visualizes profile tags as an interactive bubble chart(see Appendix \ref{app:ui_details} for visualized snapshots), where users directly adjust node salience. For tag $k$,
\begin{equation}
	I_{final}(k)=\text{clip}(I_{base}(k)+\Delta(k),0,1).
\end{equation}
To avoid permanent overfitting by transient mood shifts, manual bias decays over time:
\begin{equation}
	\Delta_{t+1}=\gamma\cdot\Delta_t,\quad \gamma=0.65.
\end{equation}

\vspace{1mm}\noindent\textbf{Backend semantic rule graph with personalized PageRank.}
The backend maintains a directed rule association graph $G_{rule}=(V,E)$ based on semantic similarity (MiniLM cosine threshold $\tau_e=0.65$). Rule salience is estimated by Personalized PageRank(PPR)\cite{haveliwala2002topic}:
\begin{equation}
	PR=\alpha M^T PR+(1-\alpha)p,
\end{equation}
where $\alpha=0.85$, $M$ is transition matrix, and $p$ is personalized prior proportional to absolute rule weights. This gives robust structural memory for unseen but semantically related items, enabling the Judge Agent to prioritize latent \textit{Meta-preferences} rather than overfitting to isolated short-term feedback.


\vspace{1mm}\noindent\textbf{Bidirectional Curation (The Star Badge).}
Moving beyond mere defense, MAP-V calculates a positive preference score for non-blocked items using the Top-$k$ weighted nodes from the user's profile graph. Items exceeding the score threshold are adorned with a ``Star Badge'' (examples provided in Appendix \ref{app:ui_details}). This white-box curation effectively mitigates users' Fear Of Missing Out (FOMO), transforming the filtering tool into a trusted content discovery companion.

\vspace{1mm}\noindent\textbf{Quantitative star scoring formulation.}
For unblocked items, MAP-V computes alignment with top-$k$ profile nodes:
\begin{equation}
\text{raw}(t)=\frac{\sum_{i=1}^{k}\pi_i\,\cos(\mathbf{e}(t),\mathbf{e}(n_i))}{\sum_{i=1}^{k}\pi_i},
\quad
s(t)=\mathrm{clip}\!\left(\frac{\text{raw}(t)-0.10}{0.40},0,1\right),
\end{equation}
where $\pi_i$ is node importance and $\mathbf{e}(\cdot)$ denotes MiniLM embedding. In our implementation, one star is shown when $s(t)\ge0.40$ and two stars when $s(t)\ge0.65$, enabling positive recommendation cues without weakening filtering constraints.

\section{Offline Adversarial Evaluation}

To assess MAP-V's upper-bound capability in suppressing inferential hallucinations and bridging the modal gap, we constructed an offline adversarial benchmark. To prevent benign content from diluting evaluation metrics, this dataset was specifically designed to push the system to its operational limits.

\subsection{Datasets and Baselines}
We curated 473 high-confusion samples from \textit{Xiaohongshu} and \textit{Zhihu} (major Chinese lifestyle and knowledge-sharing platforms) into three adversarial personas: \textbf{Persona A (Conceptual Confusion)} for contextual ambiguity; \textbf{Persona B (Emotional Nuance)} for intent differentiation; and \textbf{Persona C (Image-Text Mismatch)} for hidden visual violations. Two experts performed double-blind annotation ($0$=Pass, $1$=Block), achieving $78.86\%$ absolute agreement. While the Cohen's Kappa\cite{cohen1960coefficient} ($\kappa=0.303$) reflects the high subjectivity inherent in multi-modal filtering, all $100$ cases of disagreement were resolved through consensus to ensure a robust ground truth.

We compare MAP-V against \textbf{Baseline 1 (Keyword Matching)} and \textbf{Baseline 2 (Text-only LLM)}, which reproduces the exact text-filtering pipeline and prompt logic of the prior state-of-the-art \cite{liu2025filtering}.

\begin{table*}[t]
\caption{Overall Performance and Ablation Analysis on the Offline Adversarial Dataset ($N=473$). MAP-V demonstrates superior performance, while ablation variants reveal the orthogonal contributions of multimodal perception (Recall) and multi-agent decoupling (Precision).}
\label{tab:offline_eval}
\vspace{-2mm}
\centering
\resizebox{\textwidth}{!}{%
\begin{tabular}{l|cccc|ccc}
\toprule
\textbf{System Configuration} & \textbf{TP} & \textbf{FP (Misclass.)} $\downarrow$ & \textbf{TN} & \textbf{FN (Missed)} $\downarrow$ & \textbf{Precision} $\uparrow$ & \textbf{Recall} $\uparrow$ & \textbf{F1-Score} $\uparrow$ \\
\midrule
Baseline 1: Keyword Matching & 19 & 63 & 318 & 73 & 0.2317 & 0.2065 & 0.2184 \\
Baseline 2: Text-only LLM \cite{liu2025filtering} & 68 & 202 & 179 & 24 & 0.2519 & 0.7391 & 0.3757 \\
\midrule
Ablation A: RemoveMA (Image+Text, Single Agent) & 80 & 324 & 57 & 12 & 0.1980 & \textbf{0.8696} & 0.3226 \\
Ablation B: RemoveImage (MAP-V w/o Image) & 13 & 6 & 375 & 79 & \textbf{0.6842} & 0.1413 & 0.2342 \\
\midrule
\textbf{MAP-V (Full Pipeline)} & \textbf{80} & \textbf{52} & \textbf{329} & \textbf{12} & 0.6061 & \textbf{0.8696} & \textbf{0.7143} \\
\bottomrule
\end{tabular}%
}
\vspace{-1mm}
\end{table*}

\subsection{Performance and Ablation Analysis}
Table \ref{tab:offline_eval} details the comprehensive results integrating both the main benchmark and orthogonal ablations.

\vspace{1mm}\noindent\textbf{Eradicating Catastrophic Over-Association.} 
MAP-V significantly constrains LLM hallucinations. Baseline 2 (monolithic text LLM) generated an unacceptable 202 False Positives (FP) with a 0.2519 Precision---meaning 3 out of 4 blocked items are misclassifications, which would rapidly destroy user trust. In contrast, MAP-V slashed FPs by 74.3\% (down to 52) and elevated Precision to 0.6061. Notably, in high-noise environments like Persona B (detailed breakdown in Appendix \ref{app:persona_breakdown}), MAP-V doubled the F1-Score (0.3472 to 0.7586), proving its robustness in capturing deep-seated intents over superficial trigger words. Overall, MAP-V achieves an F1-Score of 0.7143, nearly doubling the prior SOTA.

\vspace{1mm}\noindent\textbf{Ablation: Multimodality as the ``Recall Engine.''} 
\textit{RemoveImage} strips the Qwen-VL stage but retains multi-agent logic. While Precision peaked (0.6842), Recall crashed catastrophically from 0.8696 to 0.1413, with False Negatives (FN) surging nearly sevenfold (12 to 79). This proves that without visual grounding, text-only systems are defenseless against ``image-text mismatches.'' Multimodal perception is thus the indispensable \textit{Recall Engine} raising the detection ceiling.

\vspace{1mm}\noindent\textbf{Ablation: Multi-Agent as the ``Precision Valve.''} 
\textit{RemoveMA} feeds multimodal features into a monolithic LLM, abandoning agentic decoupling. While Recall remained high (0.8696), FPs exploded to 324 (+122 vs. text baseline), plunging the F1-Score to 0.3226. Overwhelmed by rich visual data, the monolithic judge engaged in divergent speculation. This confirms that naively appending vision without architectural decoupling severely exacerbates hallucinations.

\vspace{1mm}\noindent\textbf{The Synergistic Effect.} 
Orthogonal ablations reveal a $1+1>2$ synergy: multimodality acts as the \textit{Recall Engine} discovering hidden violations, while multi-agent decoupling serves as the \textit{Precision Valve} suppressing hallucinated FPs. Isolated, each faces inherent bottlenecks; combined, they successfully decrypt high-confusion filtering.

\subsection{Qualitative Failure Cases}
Despite significant gains, MAP-V yielded 52 FPs and 12 FNs. Thematic reviews indicate FPs predominantly occur when highly aggressive title trigger words overpower benign visual evidence. Conversely, FNs typically arise from implicit visual marketing subtly embedded in everyday vlogs, challenging the VLM's zero-shot reasoning. A detailed qualitative diagnosis of these boundary cases is provided in Appendix \ref{app:failure_cases}.

\section{In-the-Wild Longitudinal Study}

While offline adversarial evaluations establish the static boundaries of an algorithm, recommendation filtering is inherently a dynamic Human-AI Co-governance process\cite{amershi2014power}. To investigate MAP-V's real-world intent alignment, governance efficiency, and impact on user perception, we deployed the system in a 7-day longitudinal in-the-wild study.


\begin{figure}[t]
  \centering
  \includegraphics[width=\linewidth]{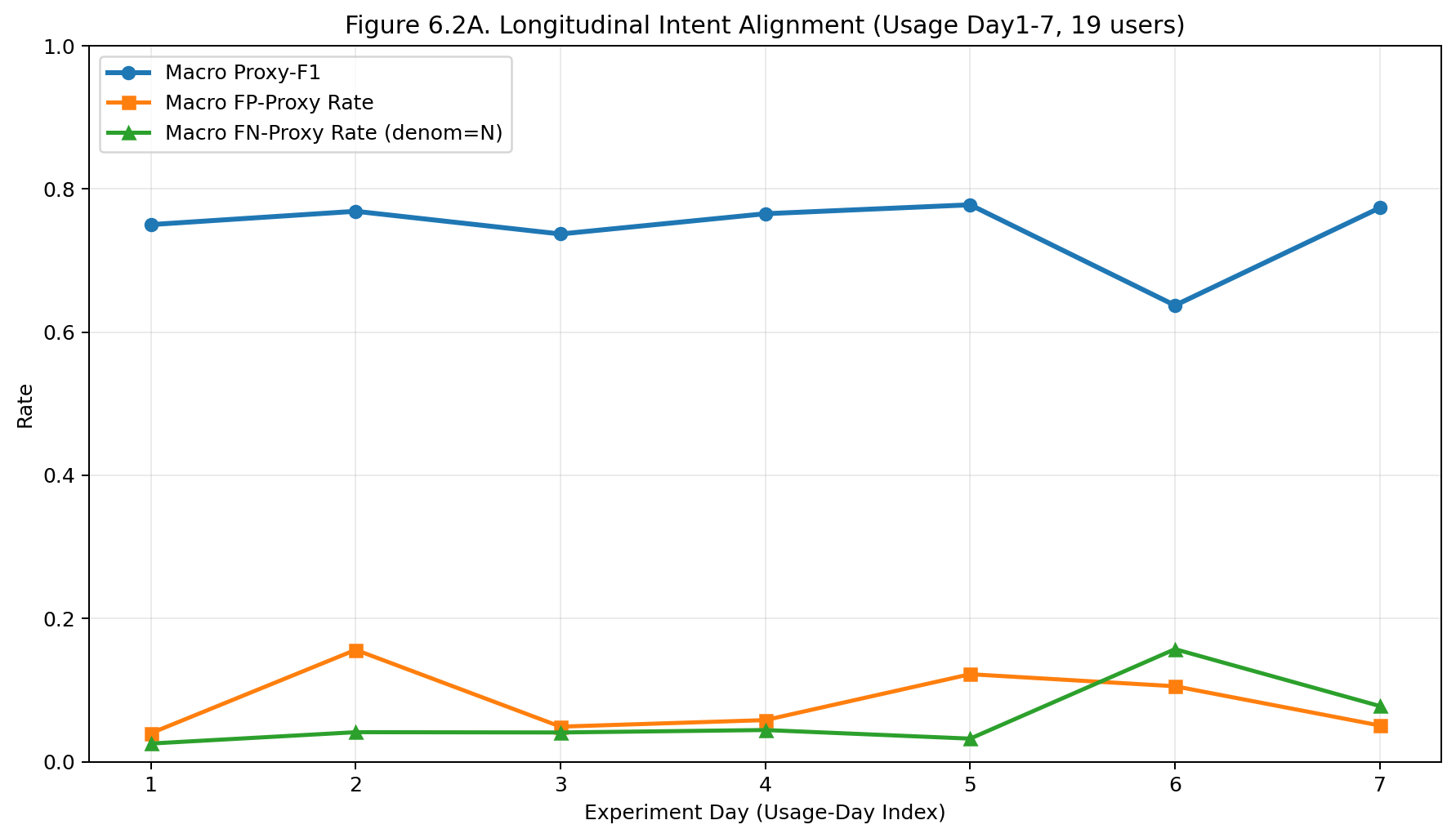}
  \vspace{-6mm}
  \caption{Longitudinal Intent Alignment (Usage Day 1-7). The system maintains a consistently high Proxy-F1 score while keeping FP and FN proxy rates minimal.}
  \label{fig:alignment}
  \vspace{-3mm}
\end{figure}

\begin{figure}[t]
  \centering
  \includegraphics[width=\linewidth]{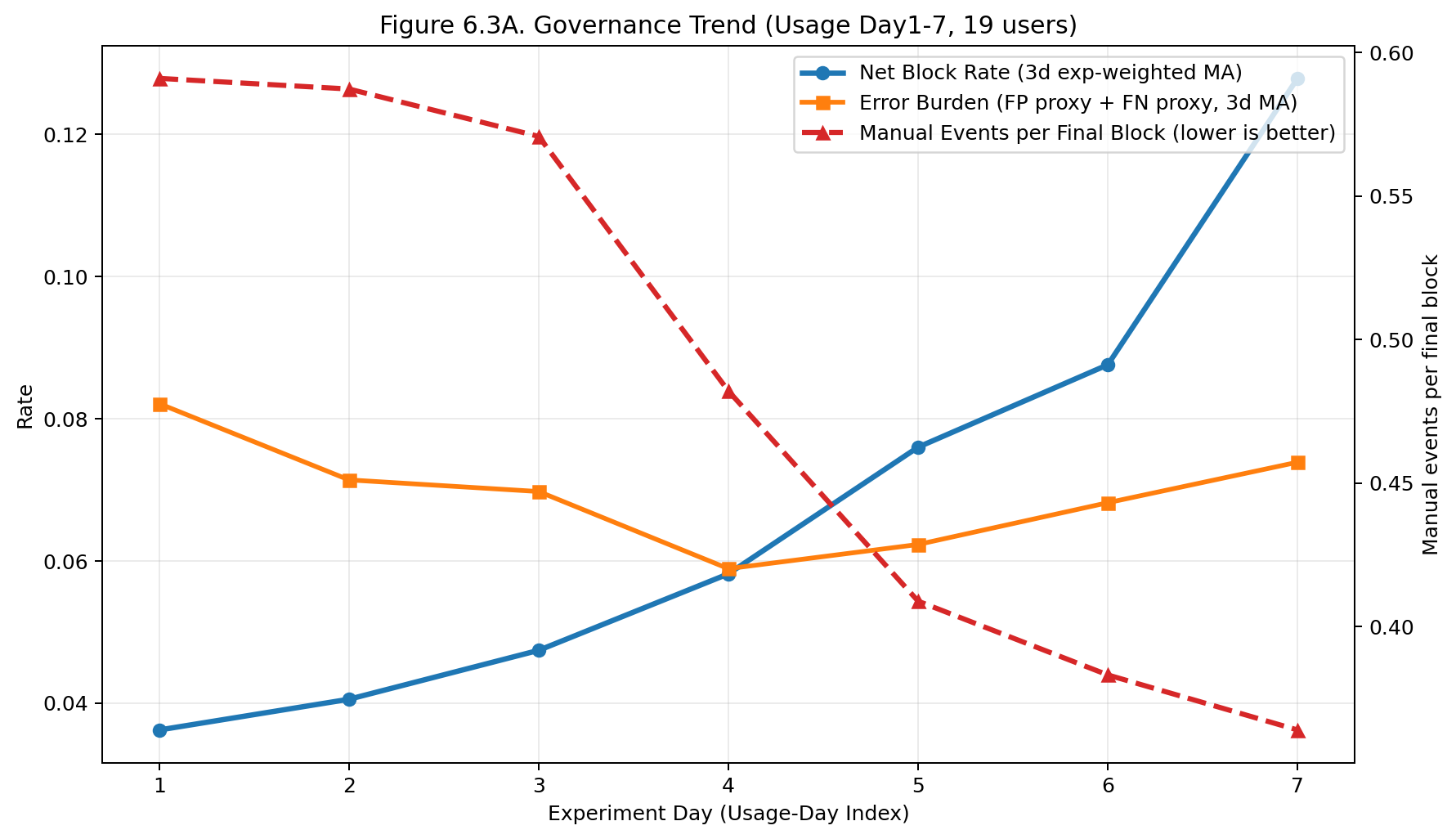}
  \vspace{-6mm}
  \caption{Governance Efficiency Trend. As the net interception rate increases, the manual events required per final block steadily decline, indicating a reduction in human effort.}
  \label{fig:governance}
  \vspace{-3mm}
\end{figure}

\subsection{Study Design and Proxy Metrics Formulation}
We recruited 19 participants who installed the MAP-V extension and naturally interacted with their daily recommendation feeds (Zhihu and Xiaohongshu). To ensure fairness despite natural usage interruptions, we aligned the timeline by ``Usage Days'' (Days 1--7), collecting a total of 66,603 valid content exposures. Participants were compensated with \$14 USD (100 RMB) upon completion.

To conduct quantitative analysis without disrupting users' natural browsing behaviors (i.e., avoiding transforming users into ``data annotators''), we formulated a rigorous \textit{Proxy Metrics} system grounded in implicit feedback research \cite{kelly2003implicit}. We strictly defined a user clicking ``Appeal/Undo'' as a \textbf{False Positive Proxy (FP Proxy)}, and a user manually ``Adding a new filter'' as a \textbf{False Negative Proxy (FN Proxy)}. Automatic blocks that remained unappealed were treated as True Positives (TP). A subsequent post-study random audit confirmed that these implicit telemetry logs achieved high fidelity in mirroring true system performance, ensuring both ecological validity and privacy protection.

\subsection{Longitudinal Intent Alignment and Governance Efficiency}

\vspace{1mm}\noindent\textbf{Stable Intent Alignment in the Wild.} 
As shown in Figure \ref{fig:alignment}, MAP-V demonstrated highly stable intent alignment across the 7-day window. The overall system achieved an impressive Precision of 0.9645, Recall of 0.7259, and an F1-Score of 0.8283, with the FP and FN proxy rates remaining exceptionally low at 0.0355 and 0.0279, respectively. Although daily metrics exhibit natural fluctuations driven by the diversity of individual user contexts, topic explorations, and evolving browsing behaviors, the aggregate performance remains consistently high. This longitudinal stability validates that MAP-V effectively translates fine-grained human intents into accurate, resilient filtering boundaries, even amidst the noise and variance of real-world environments.

\vspace{1mm}\noindent\textbf{Liberation of Manual Effort (Governance Efficiency).} 
The core value of a white-box filtering system lies not only in ``blocking more'' but in doing so without exacerbating the user's cognitive load. As depicted in Figure \ref{fig:governance}, MAP-V exhibits a remarkable ``scissors difference'' effect. Comparing the first three days to the final three days, the system achieved a \textbf{134.7\% Interception Gain}. Crucially, this was accompanied by a \textbf{33.9\% Manual Cost Reduction}---the number of manual events required per final block steadily decreased from 0.59 to 0.36. This effectively resolves the ``feedback fatigue'' prevalent in native platforms, where achieving a cleaner feed typically demands endless, repetitive ``Not Interested'' clicks without reliable long-term generalization. By contrast, MAP-V establishes a sustainable synergy between increased governance output and reduced human labor, freeing users from the exhaustion of algorithmic self-rescue.

\subsection{Subjective Perception: Breaking the Black-Box}


\begin{figure}[t]
  \centering
  \includegraphics[width=\linewidth]{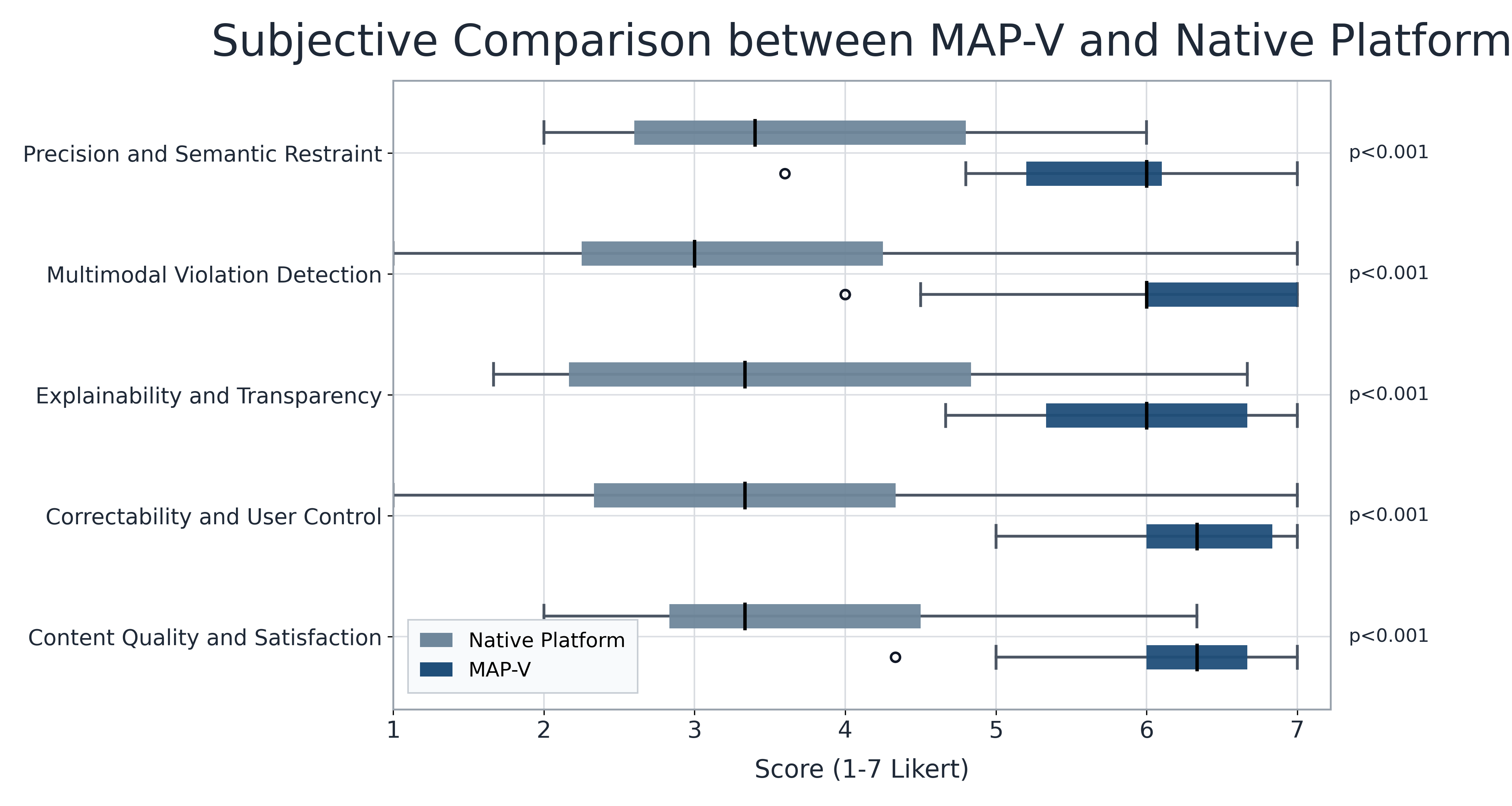}
  \vspace{-4mm}
  \caption{Subjective Comparison between MAP-V and Native Platforms across Five Core Dimensions (7-point Likert Scale). The box plots illustrate the interquartile range (IQR), median, and overall distribution. MAP-V significantly outperforms native mechanisms in all aspects ($p < 0.001$).}
  \label{fig:subjective}
  \vspace{-2mm}
\end{figure}

\begin{figure}[t]
  \centering
  \includegraphics[width=\linewidth]{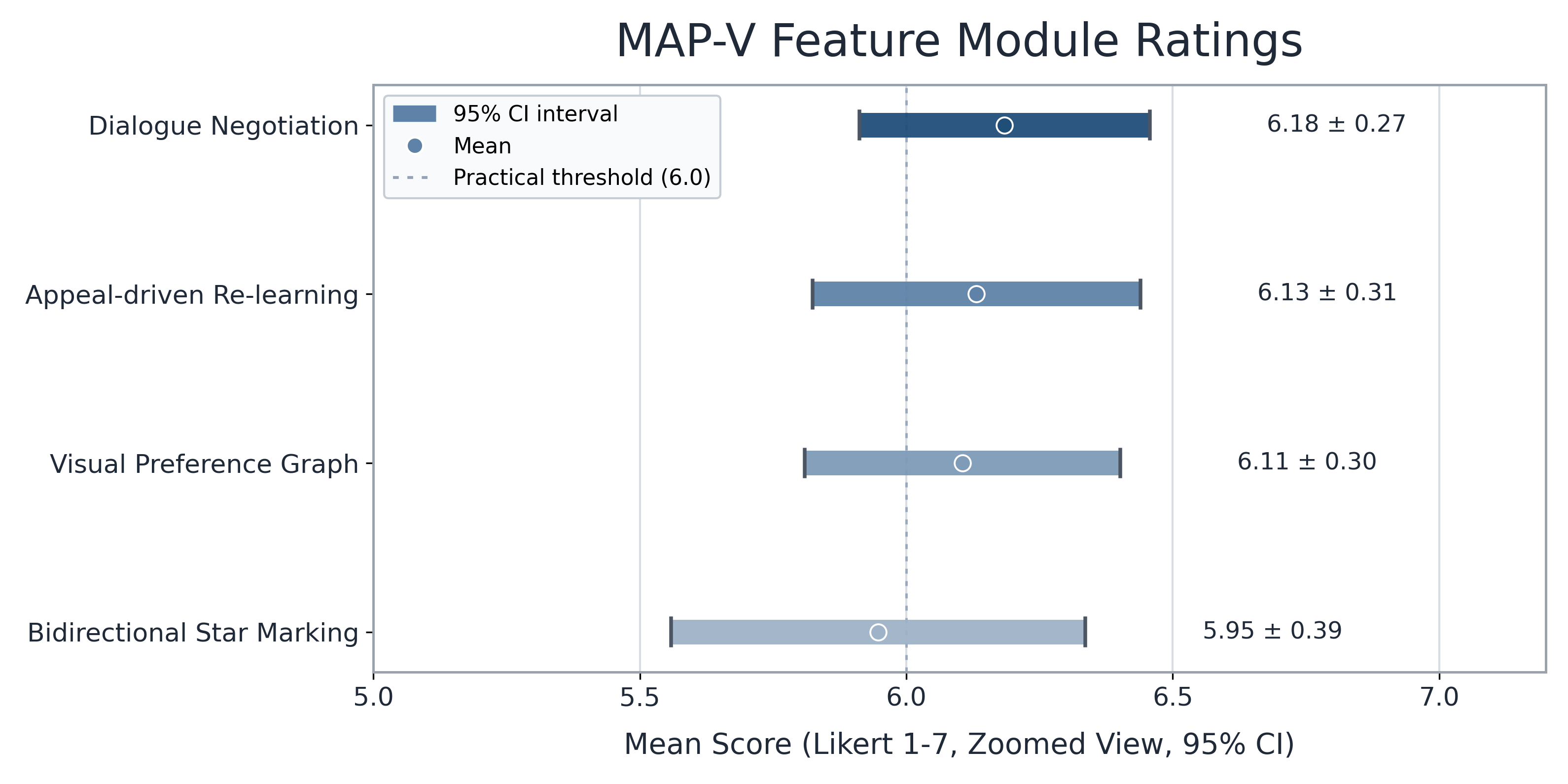}
  \vspace{-4mm}
  \caption{Usability ratings of MAP-V specific feature modules (Zoomed view showing Mean $\pm$ 95\% CI). All modules comfortably surpass the practical threshold of 6.0.}
  \label{fig:module_ratings}
  \vspace{-4mm}
\end{figure}

To validate whether the objective telemetry improvements translate into user perception, we conducted a 7-point Likert scale survey evaluating five core dimensions \textbf{(the complete questionnaire is provided in Appendix \ref{app:questionnaire})}. Paired t-tests revealed that MAP-V significantly outperformed the native platforms' ``Not Interested'' mechanisms across all dimensions ($p < 0.001$, Figure \ref{fig:subjective}).

\vspace{1mm}\noindent\textbf{Significant Leap in Controllability.} Native platforms scored a median of approximately 3.5 in ``Correctability and User Control,'' exposing the helplessness users feel towards black-box algorithms. MAP-V bridged this gap, with the mean score increasing substantially to 6.32 ($\Delta = 2.84$). This visually confirms that the explicit $\Delta$-adjustments and multi-agent negotiation successfully dismantled the system's authoritarian definition of user preferences.

\vspace{1mm}\noindent\textbf{Filling the Multimodal Blind Spot.} Regarding ``Multimodal Violation Detection,'' native platforms received negative ratings (mean 3.42), reflecting their systemic failure against implicit marketing or clickbait covers. With the Qwen-VL-Plus pipeline, MAP-V's ratings densely converged in the high-score zone ($>6.0$), perfectly corroborating our offline adversarial findings.   

\vspace{1mm}\noindent\textbf{High Usability of Agentic Features.} 
Beyond end-to-end comparisons, we conducted independent evaluations of MAP-V's frontend modules to assess the interaction friction. As shown in Figure \ref{fig:module_ratings}, introducing a human-AI negotiation workflow did not compromise system usability. The means and 95\% confidence intervals (CIs) of all modules comfortably surpassed (or closely approached) the exceptional practical threshold of 6.0. 

Specifically, \textbf{Dialogue Negotiation (6.18)} and \textbf{Appeal-driven Re-learning (6.13)} emerged as the top-rated features. These peak scores validate our hypothesis: compared to forcing users to manually write arcane regular expressions, the LLM-driven ``AI Proposal + Human One-Click Approval/Rejection'' paradigm drastically lowers the configuration barrier. Furthermore, the \textbf{Visual Preference Graph (6.11)}, characterized by its narrow CI, demonstrates that the vast majority of users felt reassured by the ability to clearly inspect and manipulate their own ``digital behavioral slices.''
\vspace{-2mm}
\subsection{Thematic Analysis and Trust Rebuilding}
To complement the quantitative data, we conducted a thematic analysis\cite{braun2006using} on open-ended feedback (yielding 99 positive descriptors); additional representative quotes are provided in Appendix~\ref{app:thematic}). Users highlighted three profound shifts in their mental models:

\vspace{1mm}\noindent\textbf{Theme 1: Multimodal Awe (Transcending Textual Deception).} Compared to traditional keyword filters, users highly appreciated MAP-V's fact-grounded visual judgment. As \textit{User09} noted: \textit{``Some posts on Xiaohongshu have seemingly normal text but borderline covers; the native platform pushes them anyway, but the plugin detects the image and blocks them effectively.''} This qualitative insight cements the indispensability of the visual layer.

\vspace{1mm}\noindent\textbf{Theme 2: FOMO Cured.} Under black-box recommendations, frequent clicking of ``Not Interested'' risks a catastrophic generalization that kills high-quality related content. MAP-V's dual-layer adjudication effectively cured this Fear Of Missing Out. \textit{User01} stated: \textit{``It successfully blocked anxiety-inducing propaganda while preserving genuine experience-sharing. It precisely differentiates complex contexts.''}

\vspace{1mm}\noindent\textbf{Theme 3: Trust Rebuilding (From Helplessness to Reason).} Over 84\% of users reported a strong sense of empowerment. \textit{User07} summarized: \textit{``It changed my distrust of the black box. Because I can provide input when blocking, making the AI understand my intent... I can actually view and modify the rules.''} The unidirectional downvote has been transformed into an explicit, transparent negotiation.

\section{Discussion}
\balance
\subsection{The Necessity of Multimodality and Factual Grounding}
Findings from our offline adversarial benchmarks ($N=473$) and 7-day longitudinal deployment underscore a critical consensus: in modern recommendation filtering, multimodal perception is not an accessory, but a fundamental prerequisite. Unimodal text models systematically fail to detect ``image-text mismatch'' violations, yet unconstrained visual augmentation exacerbates hallucinatory risks\cite{li2023evaluating}. Our ablation studies confirm that integrating Vision-Language Models with strict ``factual grounding'' protocols (Appendix~\ref{app:system_prompts}) establishes robust boundaries for LLM adjudication. By restricting the multi-agent pipeline to verifiable multimodal facts rather than speculative inference, MAP-V effectively circumvents the over-association prevalent in conventional filtering mechanisms, driving our primary gains in precision and recall.

\subsection{Engineering Trade-offs in the Dual-Graph Architecture}
A central contribution of MAP-V is balancing immediate user agency with algorithmic generalization via the Dual-Layer Preference Graph. The frontend employs a linear base-score mechanism coupled with explicit manual $\Delta$-adjustments. This design prioritizes \textit{immediate responsiveness} and \textit{linear interpretability} to alleviate early-stage trust deficits during system adoption (e.g., Usage Day 2). Conversely, the backend leverages MiniLM embeddings and PageRank for semantic deduplication and long-tail rule propagation. This architecture acknowledges that while complex graphs excel at implicit long-term profiling, human-AI trust necessitates explicit and deterministic feedback loops\cite{jannach2016user}. Together, these components formulate an interpretable defense mechanism, sustaining a 134.7\% interception gain while reducing manual effort by 33.9\%.

\subsection{Systemic Resilience and Rule Interpretability} 
Beyond predictive accuracy, deployment telemetry demonstrates MAP-V's systemic resilience and operational interpretability. 

\vspace{1mm}\noindent\textbf{Graceful Degradation via Layered Filtering.} 
Analysis of \texttt{filter\_layer} logs (Appendix \ref{app:sys_logs}) indicates that the primary cloud LLM processed 91.79\% of exposures, generating 95.85\% of initial blocks (4,891 of 5,103). The local \texttt{clip\_fallback} microservice served as a vital failsafe. Triggered by API timeouts on merely 0.32\% of exposures, it still contributed 4.15\% of total blocks. This fallback branch employs a maximally conservative 100\% block rate to ensure zero protection downtime during network latency spikes, albeit yielding marginally higher appeal rates.

\vspace{1mm}\noindent\textbf{Long-Tail Preferences and Auditable Debugging.} 
Across the 7-day window, MAP-V executed 5,103 blocks driven by 746 distinct user-generated rules. The distribution exhibits a severe long-tail effect\cite{park2008long} (Appendix \ref{app:sys_logs}): the Top 20 rules caused 31.6\% of blocks, whereas 53.7\% of rules triggered $\le 2$ times. This empirically validates the highly idiosyncratic nature of user discomfort, rendering uniform platform-level tags fundamentally inadequate. Moreover, user appeals (False Positives) remained highly concentrated; \texttt{rule\_e24f4ca1} alone generated 42.86\% of appeals. This concentration transforms abstract LLM hallucinations into \textit{auditable artifacts}, offering maintainers high-ROI (Return on Investment) pathways to surgically refine problematic boundaries without compromising long-tail governance.

\subsection{Limitations and Future Work}
Despite robust outcomes, this study presents certain limitations. First, the in-the-wild deployment ($N=19$, 7 days) exhibited a long-tail distribution in exposure volumes across participants. Although we observed strong human-AI alignment and a progressive decline in daily filtering rates (the ``Curator Effect''), we cannot establish causal evidence that the host platform's black-box algorithm experienced fundamental weight decay driven by algorithmic confounding loops \cite{chaney2018how}. Future work necessitates large-scale, isolated control groups to rigorously quantify this delayed feedback dynamic. Second, the reliance on cloud-edge collaboration and proprietary VLM APIs introduces latency vulnerabilities and privacy concerns regarding sensitive user profiles. Transitioning the multi-agent pipeline to fully localized, quantized on-device Vision-Language Models (e.g., Qwen2-VL-7B-Local)\cite{xu2024ondevice} constitutes a critical trajectory for decentralized, privacy-preserving recommendation governance.
\section{Conclusion}
The evolution of personalized recommender systems requires a paradigm shift beyond engagement-centric optimization toward controllable environments that preserve user agency. In this work, we presented MAP-V, a multimodal multi-agent framework designed to mitigate the systemic over-association and modal blindness inherent in existing LLM-based filtering solutions. By integrating a fact-grounded adjudication pipeline with a $\Delta$-adjusted dual-layer preference graph, MAP-V successfully addressed the semantic gap between high-level user complaints and low-level visual-textual features. Our comprehensive evaluation—comprising an offline adversarial benchmark and a 7-day in-the-wild longitudinal study—demonstrated that MAP-V reduces false positives by 74.3\% and decreases human governance effort by 33.9\%, without compromising filtering recall. These findings suggest that by decoupling intent parsing from factual verification and enabling explicit manual intervention, recommendation ecosystems can transition from opaque, uncontrollable black boxes into transparent, human-AI co-governed digital environments.
\begin{acks}
The authors are deeply grateful to Prof. Tun Lu for his visionary guidance in experimental design and manuscript structuring. We thank Jiahao Liu for his technical support in baseline reproduction and evaluation. We also thank the College of Computer Science and Artificial Intelligence for providing the research platform and resources.

\textbf{Author Contributions:} Both authors contributed equally. 

\textbf{C.Z.} spearheaded the system architecture and led the code refactoring and optimization of the core pipeline; he also managed data acquisition and conducted visual-modality ablation studies.

\textbf{Z.X.} led the technical implementation of high-level interaction modules and the multi-terminal deployment for multi-user experiments, as well as baseline optimization and architectural ablation.

Both authors performed data analysis and user study coordination.
\end{acks}
\bibliographystyle{ACM-Reference-Format}
\bibliography{sample-sigconf}

\appendix

\section{Granular Performance Breakdown}
\label{app:persona_breakdown}

To further validate the robustness of the MAP-V architecture against specific types of inferential hallucinations, we disaggregate the overall offline evaluation results ($N=473$) into three distinct adversarial personas: Conceptual Confusion (Persona A), Emotional Nuance (Persona B), and Image-Text Mismatch (Persona C). 

As shown in Table \ref{tab:persona_breakdown}, MAP-V consistently outperforms both the Keyword Matching (Baseline 1) and Text-only LLM (Baseline 2) approaches across all three highly ambiguous scenarios. Notably, the multi-agent adjudication significantly suppresses False Positives in Persona A and B, effectively resolving the catastrophic over-association inherent in monolithic LLM filtering.

\begin{table*}[h]
\caption{Detailed Performance Breakdown across Adversarial Personas.}
\label{tab:persona_breakdown}
\centering
\begin{tabular}{|l|cccc|ccc|}
\hline
\textbf{System} & \textbf{TP} & \textbf{FP} $\downarrow$ & \textbf{TN} & \textbf{FN} $\downarrow$ & \textbf{Precision} $\uparrow$ & \textbf{Recall} $\uparrow$ & \textbf{F1} $\uparrow$ \\
\hline
\multicolumn{8}{|l|}{\textbf{Persona A: Conceptual Confusion (N=266)}} \\
\hline
Baseline 1 (Keyword) & 6 & 26 & 201 & 33 & 0.1875 & 0.1538 & 0.1690 \\
Baseline 2 (Text-only LLM) & 32 & 109 & 118 & 7 & 0.2270 & 0.8205 & 0.3556 \\
\rowcolor{gray!10}
\textbf{MAP-V (Ours)} & \textbf{34} & \textbf{29} & \textbf{198} & \textbf{5} & \textbf{0.5397} & \textbf{0.8718} & \textbf{0.6667} \\
\hline
\multicolumn{8}{|l|}{\textbf{Persona B: Emotional Nuance (N=169)}} \\
\hline
Baseline 1 (Keyword) & 9 & 31 & 100 & 29 & 0.2250 & 0.2368 & 0.2308 \\
Baseline 2 (Text-only LLM) & 25 & 81 & 50 & 13 & 0.2358 & 0.6579 & 0.3472 \\
\rowcolor{gray!10}
\textbf{MAP-V (Ours)} & \textbf{33} & \textbf{16} & \textbf{115} & \textbf{5} & \textbf{0.6735} & \textbf{0.8684} & \textbf{0.7586} \\
\hline
\multicolumn{8}{|l|}{\textbf{Persona C: Image-Text Mismatch (N=38)}} \\
\hline
Baseline 1 (Keyword) & 4 & 6 & 17 & 11 & 0.4000 & 0.2667 & 0.3200 \\
Baseline 2 (Text-only LLM) & 11 & 12 & 11 & 4 & 0.4783 & 0.7333 & 0.5789 \\
\rowcolor{gray!10}
\textbf{MAP-V (Ours)} & \textbf{13} & \textbf{7} & \textbf{16} & \textbf{2} & \textbf{0.6500} & \textbf{0.8667} & \textbf{0.7429} \\
\hline
\end{tabular}
\end{table*}

\section{System Prompts Formulation}
\label{app:system_prompts}

\textit{Note: The original system prompts were implemented in Chinese to maximize the Large Language Model's reasoning performance and instruction adherence. Following established prompt engineering patterns (e.g., persona adoption and format enforcement) \cite{white2023prompt}, the prompts presented below have been accurately translated into English and reformatted into a structured JSON representation for the convenience of reviewers.}

\lstset{
  basicstyle=\ttfamily\small,
  breaklines=true,
  captionpos=b,
  frame=none, 
  aboveskip=1em,
  belowskip=1em,
  showstringspaces=false
}

\subsection{Visual Feature Extraction Prompt}
\textbf{Agent Functionality:} Analyzes a raw image in the cloud and produces a structured general-purpose feature representation. The prompt restricts the extraction task to a strict three-layer structure and prohibits moral guessing.

\begin{lstlisting}[caption={Prompt Structure for Visual Feature Extraction Agent}]
{
  "role": "You are an image feature extraction engine. Objectively parse the input image and output a standardized three-layer JSON for content filtering adjudication.",
  "three_layer_framework": {
    "layer_1_perception": "Low-level visual signals: image_quality, brightness, color_temperature, composition.",
    "layer_2_cognition": "Mid-level entity recognition: subjects, demographics, appearance, object_details, actions, ocr.",
    "layer_3_semantics": "High-level content inference: scene, style, vibe, category."
  },
  "special_cases": [
    "Multi-image/Collage: Set composition to 'Nine-grid collage' or 'Multi-image collage'. Summarize subjects and category for the whole image. Extract only key info for OCR.",
    "Screenshot (Text/Code/Bill): Set style to 'Text screenshot', focus on OCR, other fields can be abbreviated.",
    "If a field cannot be determined, fill null. Guessing is prohibited."
  ],
  "output_requirements": "Only output a single line of compact JSON."
}
\end{lstlisting}

\subsection{Rule Judge Prompt}
\textbf{Agent Functionality:} The LLM serves as the content moderation judge, receiving the standardized output of visual features and text data, and logically comparing them against the user's rules.

\begin{lstlisting}[caption={Prompt Structure for Rule Adjudication Agent}]
{
  "role": "You are a content moderation judge. Based on the filtering rules set by the user, judge whether the current recommended content should be blocked.",
  "rule_intensity_description": {
    "Strong_Filter": "Triggers if the content is clearly related to the rule theme, without strict exact matching.",
    "Medium_Filter": "Triggers when clear related signals exist in the title or text.",
    "Mild_Filter": "Triggers only when the content highly matches the rule."
  },
  "adjudication_principles": [
    "Only cite actual text in the content and fact features annotated in the image as the basis for judgment.",
    "Strictly prohibit triggering rules via: implicit intent inference, background knowledge association, or 'the title implies XX'.",
    "If a rule can only be matched through inference, it must not trigger."
  ],
  "proper_noun_protection": "Proper nouns appearing in the title must be recognized as a whole. Its literal meaning does not constitute a basis for triggering unless the context refers strictly to the substance. Bad Example: 'Shameless' triggers 'Block derogatory content'.",
  "feed_preview_handling": "If only an information feed preview summary is available, do not block based on 'lack of substance' or 'clickbait'. Rely solely on the title, tags, and image analysis.",
  "output_format": {
    "filter_decision": "true/false",
    "triggered_rule_id": "Rule ID or null",
    "reason": "A one-sentence explanation citing actually existing text or image features, limited to 100 words. Rule IDs or implicit words like 'implies' are forbidden."
  }
}
\end{lstlisting}

\subsection{Intent Parser Prompt}
\textbf{Agent Functionality:} Acts as an expert in user intent analysis, responsible for translating the user's natural language feedback into structured and precise filtering rules tailored to different social media platforms.

\begin{lstlisting}[caption={Prompt Structure for Intent Parser Agent}]
{
  "role": "You are a recommender system user intent analysis expert. Your task is to convert the user's natural language feedback into a precise filtering rule.",
  "platform_characteristics": {
    "Zhihu": "Q&A/Article platform, text-centric. Rules focus on topic domains and content tendencies (e.g., 'emotional clickbait').",
    "Xiaohongshu": "Image-text/Short video platform. Rules must incorporate visual features (e.g., 'covers holding milk tea')."
  },
  "rule_description_precision": {
    "requirements": "Must be highly specific with clear qualifiers, localizable to topic domains, specific entities, or visual/formal features.",
    "bad_example": "'Block anime' (Too broad)",
    "good_example": "'Block anime image-text content with obvious sexually suggestive or vulgar tendencies'"
  },
  "weight_anchors": {
    "-1.0": "Extreme rejection, filters on slight relevance.",
    "-0.8": "Strong rejection, triggers on obvious signals in title or image.",
    "-0.6": "Moderate rejection, triggers when content clearly belongs to the theme.",
    "-0.4": "Mild rejection, triggers when topic highly matches the theme.",
    "-0.2": "Weak rejection, rarely triggers, use with caution."
  },
  "output_format": {
    "nl_description": "Precise filter rule description.",
    "core_entities": ["Core entity 1", "Core entity 2"],
    "weight": -0.8,
    "modality": "text / image / image_text"
  }
}
\end{lstlisting}

\section{Qualitative Failure Cases}
\label{app:failure_cases}

During the offline high-confusion adversarial evaluation phase, although MAP-V reduced the false positive (FP) rate by 74.3\%, there were still a very small number of false negatives (FN) and false positives (FP). We analyzed typical samples that led to such results.

\vspace{1mm}\noindent\textbf{Case 1: False Negative (Missed Detection)}
\begin{itemize}
    \item \textbf{User Rule:} ``Reject appearance anxiety and body involution''
    \item \textbf{Content Context:} Image shows a multi-image collage of outfit guides by an extremely thin model. Text states: \textit{``The weather is so nice today, casually sharing a few snaps\textasciitilde{}''}
    \item \textbf{MAP-V Judgment:} Not blocked. The poster used extremely colloquial and neutral daily trigger words to mask the visual pressure. 
    \item \textbf{Improvement Insight:} Visually implicit mental pressure conveyed solely by images remains challenging if the text employs semantic obfuscation.
\end{itemize}

\vspace{1mm}\noindent\textbf{Case 2: False Positive (Misclassification)}
\begin{itemize}
    \item \textbf{User Rule:} ``Block gender antagonism or flame-war related speech''
    \item \textbf{Content Context:} Image of the movie \textit{Barbie}. Text states: \textit{``I don't understand why this movie has comments discussing meaningless flame-wars.''}
    \item \textbf{MAP-V Judgment:} Wrongly blocked. The text is a criticism and sarcasm (meta-commentary) of flame-war behaviors, but high-frequency keywords triggered the fallback safety net.
    \item \textbf{Improvement Insight:} Sarcastic meta-commentary is a challenge. In contexts with strong keyword stacking, the ``factual basis'' rule may still lead to opposed judgments.
\end{itemize}

\section{Qualitative Success Cases}
\label{app:success_cases}

To gain a deeper understanding of MAP-V's improvement mechanisms and its advantages over pure-text systems, we selected typical success cases from various personas.

\vspace{1mm}\noindent\textbf{Case 3: Conceptual Confusion (Should Pass)}
\begin{itemize}
    \item \textbf{Topic:} \textit{``Is [Anonymous University] going to decline?''}
    \item \textbf{Baseline 2 (FP):} It only analyzes the title text. ``[Anonymous University]'' and ``decline'' create a semantic association with the user's ``campus negative events'' rule, leading the LLM to aggressively filter.
    \item \textbf{MAP-V (Correct TN):} Qwen-VL analyzes the illustration and finds \textit{``Exhibition board reflects a cultural theme, figures sit respectfully in ancient official robes, overall solemn and academic.''} The visual academic nature provides crucial disambiguation for the title's ambiguity.
\end{itemize}

\vspace{1mm}\noindent\textbf{Case 4: Implicit Marketing (Should Block)}
\begin{itemize}
    \item \textbf{Topic:} \textit{``Vlog | Post-00s Couple | A day of queuing for grilled fish''}
    \item \textbf{Baseline 2 (FN):} The title seems like a normal food exploration sharing. Pure text analysis cannot identify its implicit marketing intent.
    \item \textbf{MAP-V (Correct TP):} Image analysis found \textit{``non-catering core elements such as tarot cards and cat-ear thermal bags... showing implicit product placement tendencies.''} This confirmed a soft advertisement masquerading as a vlog.
\end{itemize}

\vspace{1mm}\noindent\textbf{Case 5: Image-Text Mismatch (Should Block)}
\begin{itemize}
    \item \textbf{Topic:} \textit{``Is RAG still meaningful now that LLMs have infinite context?''}
    \item \textbf{Baseline 2 \& MAP-V (TP):} Both blocked it, but MAP-V provided a more precise factual reason: \textit{``close-up anime-style female character illustration... with zero semantic relevance to the technical discussion.''} Even if the text-only system accidentally filtered correctly, MAP-V's ``image verification'' logic makes the decision auditable.
\end{itemize}

\section{MAP-V User Interface and Functional Walkthrough}
\label{app:ui_details}

\begin{figure*}[ht]
  \centering
  \includegraphics[width=0.98\textwidth]{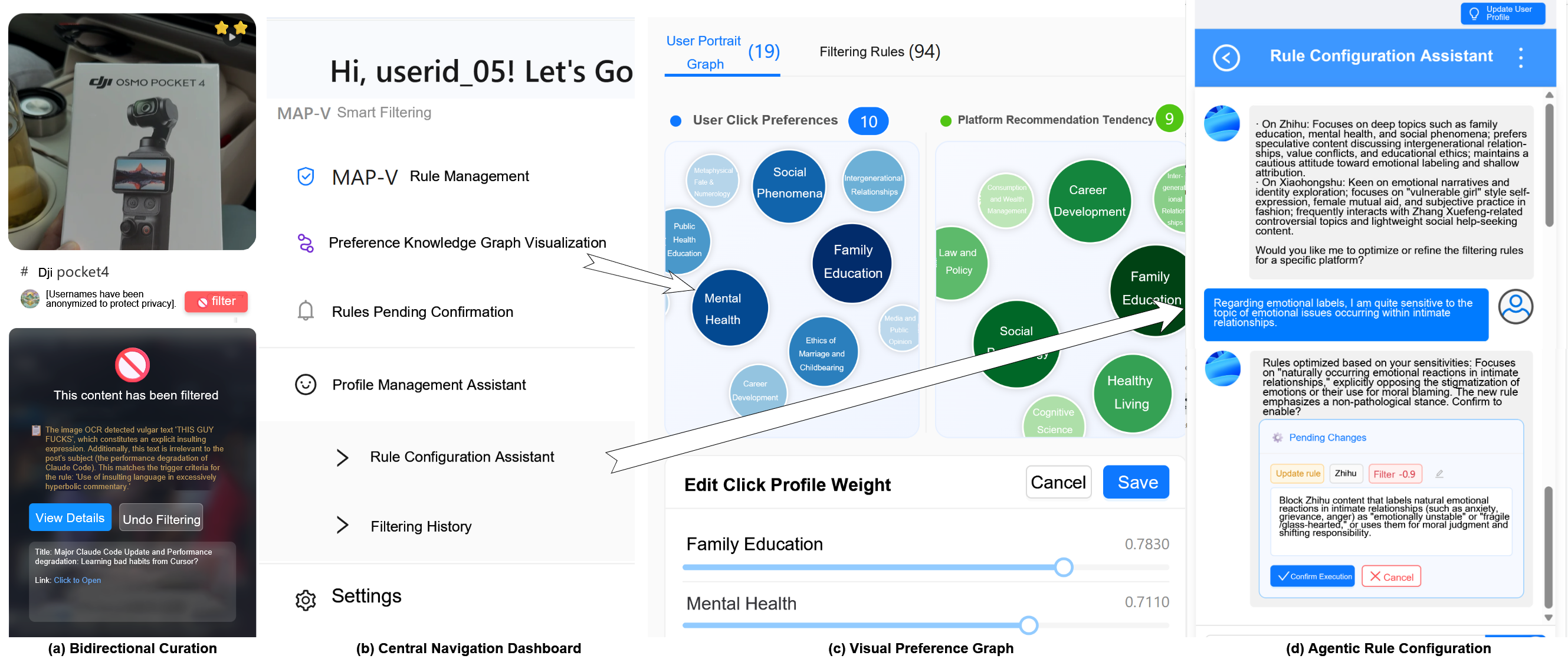}
  \vspace{-2mm}
  \caption{Detailed user interface of MAP-V. (a) \textbf{Bidirectional Curation}: Demonstrates a masked item (left) with its adjudication rationale and an unblocked item (right) adorned with a Star Badge. (b) \textbf{Central Navigation Dashboard}: The entry point for rule management, preference visualization, and agentic assistance. (c) \textbf{Visual Preference Graph}: Separates user click profiles (blue) from platform recommendation tendencies (green), allowing for manual $\Delta$-bias adjustments via sliders. (d) \textbf{Agentic Rule Configuration}: A chat-based interface where the Dispute Agent proposes rule refinements based on user feedback.}
  \label{fig:ui_appendix}
\end{figure*}

To provide a comprehensive view of the MAP-V system's interaction design, we present the full user interface in Figure~\ref{fig:ui_appendix}. The system is implemented as a cross-platform browser extension that seamlessly integrates with the DOM of recommendation platforms (e.g., Zhihu, Xiaohongshu).

\vspace{1mm}\noindent\textbf{Bidirectional Curation Mechanism (Fig. \ref{fig:ui_appendix}a):} 
When the Judge Agent identifies a violation, a semi-transparent mask is rendered over the card, providing a button to ``View Details'' for factual transparency and an ``Undo'' button for immediate correction. Simultaneously, for high-quality items, the system renders one or two star badges based on the calculated alignment score $s(t)$.

\vspace{1mm}\noindent\textbf{Interactive Profile Editing (Fig. \ref{fig:ui_appendix}c):} 
Unlike traditional black-box profiles, MAP-V exposes behavioral tags as interactive bubbles. When a user clicks a node, a slider appears (shown in the bottom of Fig. \ref{fig:ui_appendix}c), allowing the user to manually increase or decrease the importance. This action triggers a $\Delta$-bias update in the backend database.

\vspace{1mm}\noindent\textbf{Agent-Led Rule Negotiation (Fig. \ref{fig:ui_appendix}d):} 
The chatbot interface supports natural language intent parsing. As shown in the screenshot, when a user expresses a nuanced sensitivity, the agent drafts a ``Pending Change'' card, allowing the user to inspect the newly formulated rule and its associated weight before final activation.

\section{Systemic Resilience and Auditable Logs}
\label{app:sys_logs}

To substantiate the engineering resilience discussed in Section 6.3, we provide a comprehensive breakdown of the \texttt{filter\_layer} telemetry logs collected during the 7-day in-the-wild study. 

As shown in Table \ref{tab:layer_logs}, the dual-layer architecture successfully guarantees continuous protection. The primary \texttt{cloud} LLM handled 91.79\% of exposures and contributed 95.85\% of the original blocks. Crucially, when the primary cloud LLM becomes unavailable, the local \texttt{clip\_fallback} branch efficiently takes over. Although it only covered 0.32\% of exposures due to rare API timeouts, it executed a 100\% block rate as a conservative safety measure, contributing 4.15\% of the total original blocks. The \texttt{pass} branch naturally covered remaining exposures without executing any blocks, aligning with the default ``allow'' semantics.

\begin{table}[h]
\caption{Comprehensive Distribution of Filtering Layers (Usage Day 1-7). The log details confirm that the local fallback mechanism acts as a robust safety net with zero protection downtime.}
\label{tab:layer_logs}
\centering
\resizebox{\columnwidth}{!}{%
\begin{tabular}{|lrrrrrr|}
\hline
\textbf{Filter Layer} & \textbf{Exposures} & \textbf{Orig. Blocks} & \textbf{Final Blocks} & \textbf{Appeals} & \textbf{Block Rate} & \textbf{Appeal Rate} \\
\hline
\texttt{cloud} & 61,136 & 4,891 & 4,723 & 168 & 8.00\% & 3.43\% \\
\texttt{pass} & 5,234 & 0 & 0 & 0 & 0.00\% & 0.00\% \\
\texttt{clip\_fallback} & 212 & 212 & 199 & 13 & 100.00\% & 6.13\% \\
\texttt{unknown} & 21 & 0 & 0 & 0 & 0.00\% & 0.00\% \\
\hline
\end{tabular}%
}
\end{table}

\section{The Long-Tail Distribution of User Preferences}
\label{app:long_tail}

We further analyzed the 5,103 original blocks executed by MAP-V. These blocks involved 746 distinct user-generated rules. The distribution exhibits a profound long-tail effect: approximately 37.5\% of rules were triggered only once, and 53.7\% were triggered $\le 2$ times. Table \ref{tab:top_rules} displays the Top 15 most frequently triggered rules, which collectively accounted for only a fraction of total blocks, proving that the vast majority of protection relies on highly personalized, long-tail rules.

Conversely, algorithmic hallucinations (False Positives) are disproportionately concentrated in a few specific rules. As detailed in Table \ref{tab:high_appeal}, the top appealed rules reveal critical insights for system maintenance. For example, \texttt{rule\_e24f4ca1} alone triggered 18 appeals, yielding a concentrated appeal rate of 42.86\%. This auditable transparency allows developers to perform targeted, high-ROI debugging on problematic boundary rules without disrupting the personalized long-tail protection that users rely on.

\begin{table}[h]
\caption{Top 15 Most Triggered Rules by Original Blocks. The data highlights a long-tail distribution where top rules account for only a limited portion of total governance.}
\label{tab:top_rules}
\centering
\resizebox{\columnwidth}{!}{%
\begin{tabular}{|lrrrr|}
\hline
\textbf{Triggered Rule ID} & \textbf{Orig. Blocks} & \textbf{Final Blocks} & \textbf{Appeals} & \textbf{Appeal Rate} \\
\hline
\texttt{rule\_9e166a88} & 178 & 178 & 0 & 0.00\% \\
\texttt{rule\_25cee3e0} & 171 & 170 & 1 & 0.58\% \\
\texttt{rule\_c9647f0b} & 166 & 164 & 2 & 1.20\% \\
\texttt{rule\_3632c6a2} & 122 & 121 & 1 & 0.82\% \\
\texttt{rule\_8d30905d} & 110 & 110 & 0 & 0.00\% \\
\texttt{rule\_89ce876f} & 99 & 99 & 0 & 0.00\% \\
\texttt{rule\_f18a7b51} & 96 & 96 & 0 & 0.00\% \\
\texttt{rule\_55} & 69 & 69 & 0 & 0.00\% \\
\texttt{rule\_d0883794} & 67 & 63 & 4 & 5.97\% \\
\texttt{rule\_c7d0b6c6} & 60 & 59 & 1 & 1.67\% \\
\texttt{rule\_8c9639c6} & 56 & 56 & 0 & 0.00\% \\
\texttt{rule\_04087b2e} & 53 & 53 & 0 & 0.00\% \\
\texttt{rule\_331e2d67} & 51 & 51 & 0 & 0.00\% \\
\texttt{rule\_bf7f6818} & 51 & 51 & 0 & 0.00\% \\
\texttt{rule\_660c8013} & 48 & 48 & 0 & 0.00\% \\
\hline
\end{tabular}%
}
\end{table}

\begin{table}[h]
\caption{Concentration of Appeals among Triggered Rules (Top 10). Unlike the long-tail distribution of preferences, hallucinations are highly concentrated and debuggable.}
\label{tab:high_appeal}
\centering
\resizebox{\columnwidth}{!}{%
\begin{tabular}{|lrrr|}
\hline
\textbf{Triggered Rule ID} & \textbf{Orig. Blocks} & \textbf{Appeals Passed} & \textbf{Appeal Rate} \\
\hline
\texttt{rule\_e24f4ca1} & 42 & 18 & \textbf{42.86\%} \\
\texttt{rule\_eda31a0b} & 29 & 6 & 20.69\% \\
\texttt{rule\_130} & 21 & 2 & 9.52\% \\
\texttt{rule\_70} & 22 & 2 & 9.09\% \\
\texttt{rule\_b1ef5cc9} & 23 & 2 & 8.70\% \\
\texttt{rule\_d0883794} & 67 & 4 & 5.97\% \\
\texttt{rule\_de6e1172} & 23 & 1 & 4.35\% \\
\texttt{rule\_c1a3cef8} & 23 & 1 & 4.35\% \\
\texttt{rule\_6f34df37} & 26 & 1 & 3.85\% \\
\texttt{rule\_7f6438db} & 28 & 1 & 3.57\% \\
\hline
\end{tabular}%
}
\end{table}
\section{User Study Questionnaire and Scales}
\label{app:questionnaire}

To evaluate the subjective user experience, we administered a post-study questionnaire based on a 7-point Likert scale (1 = Strongly Disagree, 7 = Strongly Agree). Adapted from established user-centric evaluation frameworks for recommender systems (e.g., the ResQue framework \cite{pu2011user}), the measurement instrument was structured around five core theoretical constructs:

\begin{description}
    \item[Precision \& Semantic Restraint] \hfill \\
    \textit{Measures the system's ability to minimize over-association.}
    \begin{itemize}
        \item \textbf{Q1.} The system precisely understands my filtering intent without over-generalization.
        \item \textbf{Q2.} The system correctly distinguishes between similar keywords in different contexts (e.g., academic sharing vs. anxiety marketing).
    \end{itemize}

    \item[Multimodal Violation Detection] \hfill \\
    \textit{Evaluates the efficacy of the visual adjudication layer.}
    \begin{itemize}
        \item \textbf{Q3.} The system accurately identifies violations hidden in images (e.g., clickbait covers) that text-only systems miss.
        \item \textbf{Q4.} The system successfully verifies the factual content within images rather than relying on titles alone.
    \end{itemize}

    \item[Explainability \& Transparency] \hfill \\
    \textit{Assesses the clarity of the ``White-box'' mechanism.}
    \begin{itemize}
        \item \textbf{Q5.} The reasons provided for each block allow me to clearly understand the system's logic.
        \item \textbf{Q6.} The Visualized Preference Graph accurately reflects my digital behavioral profile.
    \end{itemize}

    \item[Correctability \& User Control] \hfill \\
    \textit{Captures the user's perceived agency over the recommendation feed.}
    \begin{itemize}
        \item \textbf{Q7.} When a misclassification occurs, I possess effective and direct means (e.g., appeal/sliders) to correct it.
        \item \textbf{Q8.} I feel I have regained agency and control over my recommendation feed.
    \end{itemize}

    \item[Content Quality \& Satisfaction] \hfill \\
    \textit{Measures the overall impact on the information diet.}
    \begin{itemize}
        \item \textbf{Q9.} The ``Star Badge'' mechanism significantly mitigates my Fear of Missing Out (FOMO) on high-quality content.
        \item \textbf{Q10.} Overall, I am more satisfied with the quality of my feed compared to the native platform.
    \end{itemize}
\end{description}

\section{Qualitative Insights: Supporting Quotes}
\label{app:thematic}

In addition to the quantitative metrics discussed in Section 6, we provide additional representative quotes from the 7-day longitudinal study to substantiate our thematic analysis:

\vspace{2mm}\noindent\textbf{Theme: Restoration of Algorithmic Agency} \\
\textit{``It changed my inherent distrust of the black box. Because I can provide natural language input when blocking, the system finally understands my intent... Being able to view and modify the rules directly makes me feel like the master of the feed, not a passive target.''} --- \textbf{Actual\_user07}

\vspace{2mm}\noindent\textbf{Theme: Multimodal Awareness} \\
\textit{``I was impressed by how it handled 'image-text mismatch.' Native platforms constantly push posts with normal text but borderline covers; the MAP-V plugin was able to detect the actual image content and block them effectively.''} --- \textbf{Actual\_user09}

\vspace{2mm}\noindent\textbf{Theme: FOMO Mitigation} \\
\textit{``The star badges are a game-changer. They help me quickly locate high-value content amidst the noise, which makes me feel much safer about being aggressive with my filtering rules.''} --- \textbf{Actual\_user03}

\vspace{2mm}\noindent\textbf{Theme: Trust through Correction} \\
\textit{``Even when it made a mistake, the appeal process felt logical. I didn't feel frustrated by the error because I had a way to 'reason' with the AI and see it improve immediately.''} --- \textbf{Actual\_user05}

\end{document}